\documentclass[%
 reprint,
superscriptaddress,
 amsmath,amssymb,
 aps,
]{revtex4-2}

\usepackage{graphicx}
\usepackage{dcolumn}
\usepackage{bm}

\begin{document}

\preprint{APS/123-QED}

\title{THz Field-induced Spin Dynamics in Ferrimagnetic Iron Garnets }
\author{T.G.H. Blank}
\affiliation{Radboud University, Institute for Molecules and Materials, 6525 AJ Nijmegen, The Netherlands.}
\author{K.A. Grishunin}
\affiliation{Radboud University, Institute for Molecules and Materials, 6525 AJ Nijmegen, The Netherlands.}
\author{E.A. Mashkovich}
\affiliation{Radboud University, Institute for Molecules and Materials, 6525 AJ Nijmegen, The Netherlands.}
\author{M.V. Logunov}
\affiliation{Kotel'nikov Institute of Radioengineering and Electronics, 125009 Moscow, Russia.}
\author{A.K. Zvezdin}
\affiliation{Prokhorov General Physics Institute of the Russian Academy of Sciences, 119991 Moscow, Russia.}
\author{A.V. Kimel}
\affiliation{Radboud University, Institute for Molecules and Materials, 6525 AJ Nijmegen, The Netherlands.}

\date{\today}

\begin{abstract}
THz magnetization dynamics is excited in ferrimagnetic thulium iron garnet with a picosecond, single-cycle magnetic field pulse and seen as a high-frequency modulation of the magneto-optical Faraday effect. Data analysis combined with numerical modelling and evaluation of the effective Lagrangian allow us to conclude that the dynamics corresponds to the exchange mode excited by Zeeman interaction of the THz field with the antiferromagnetically coupled spins. We argue that THz-pump | IR-probe experiments on ferrimagnets offer a unique tool for quantitative studies of dynamics and mechanisms to control antiferromagnetically coupled spins. 
\end{abstract}

\maketitle

All magnetically ordered materials, depending on the alignment of spins, are divided into two primary classes: ferro- and antiferromagnets. Ferromagnets are characterized by parallel alignment of spins which results in net magnetic moment, while spins in antiferromagnets are aligned in a mutually antiparallel way with zero net magnetization in the unperturbed state. Antiferromagnets represent the largest, but the least explored class of magnets with a potential to have a dramatic impact on spintronics and other magnetic technologies. In particular, the higher frequency ($\sim$ THz) of spin resonances in antiferromagnets can bring the clock-speed of spintronics devices into the THz range \cite{RevModPhys.90.015005, NatKimelAFMspintronics, AfmspintronicsNanoNat}. 

Unfortunately, proceedings in both fundamental research and the development of antiferromagnetic spintronics are considerably hindered by the lack of net magnetization in antiferromagnets, as even the discovery of antiferromagnetic order itself had to wait for the advent of neutron diffraction experiments in the late 1940s \cite{PhysRev.76.1256.2}. This is why approaches and mechanisms allowing efficient excitation of antiferromagnetic spins in the THz range became a subject of not only intense, but also challenging and intriguing research. In particular, recently it was suggested that THz magnetic fields can excite antiferromagnetically coupled spins with a significantly higher efficiency when accounting for the new, relativistic mechanism of field derivative torque (rFDT) \cite{Mondal2016}. This torque can reach strengths comparable with conventional the Zeeman torque \cite{Mondal2019}. However, the lack of methods for quantitative detection of spins in antiferromagnets prevents these claims from experimental verification and can even lead to mistakes in interpretation of experimental results \cite{PhysRevLett.124.039901}. 

A substantial progress in understanding THz light-spin coupling can be achieved by studying ferrimagnets, which are a subclass of antiferromagnets having two non-equivalent magnetic sublattices. Within each sublattice the spins are aligned ferromagnetically, while the intersublattice interaction is antiferromagnetic. The sublattice magnetizations can be different in size, and therefore the net magnetization is not necessarily zero. The latter greatly simplifies experimental studies, but it does not ruin the presence of THz resonances called exchange modes, as antiferromagnetic order is still present. In this article, we demonstrate and explore the high-frequency response of antiferromagnetic spins in a ferrimagnet to THz magnetic field. We experimentally reveal the orientation of the THz field which causes the largest deviation of spins from their equilibrium. Using simulations we show that the oscillations correspond to the exchange mode of spin resonance. The applied experimental technique is shown to have a great potential to facilitate quantitative conclusions. In particular, due to the non-zero Faraday rotation in the unperturbed state ($\alpha_F$) and having the calibrated dynamic Faraday rotation ($\Delta \alpha_F$), the ratio $\Delta \alpha_F / \alpha_F$ unambiguously defines spin deviations caused by the calibrated THz magnetic field. The technique allows us to show that the conventional Zeeman torque does play in the spin-excitation the dominant role, while alternative mechanisms can essentially be neglected.

The garnet structure (crystallographic space group Ia$\bar{3}$d) of rare-earth iron garnets (REIGs) gives rise to unusual magnetic properties \cite{neelferri, HoIG_spinflop}. Three of five Fe\textsuperscript{3+}-ions per formula unit (R\textsubscript{3}Fe\textsubscript{5}O\textsubscript{12}) form a sublattice with tetrahedral symmetry and are antiferromagnetically coupled to the remaining two iron ions occupying sites of octahedral symmetry. The imbalance between these iron ions results in a net magnetic moment $\mathbf{M}_{Fe}$ to which the rare-earth site magnetization $\mathbf{M}_{R}$ aligns anti-parallel. The result is a three-sublattice ferrimagnet with net magnetization $\mathbf{M} = \mathbf{M}_{R} + \mathbf{M}_{Fe}$. The antiferromagnetic exchange between the iron sublattices is large compared to any other interactions experienced by the Fe\textsuperscript{3+} spins, justifying the approximation of treating it as a single sublattice with magnetization $\mathbf{M}_{Fe}$ \cite{Levitin}. The RE-sublattice experiences the exchange-field generated by this iron magnetization \cite{neelferri}, while intra-sublattice exchange interaction is weak and can be ignored, resembling a paramagnet in the exchange field. 

The REIG structure studied in this work is a $19$ $\mu$m film of Bi- and Ga- substituted thulium iron garnet Tm\textsubscript{3-x}Bi\textsubscript{x}Fe\textsubscript{5-y}Ga\textsubscript{y}O\textsubscript{12} (TmIG) with targeted composition $x = 1$, $y = 0.8$. The film was grown by liquid phase epitaxy on a $500$ $\mu$m thick ($111$)-oriented GGG substrate. The sample was doped with Bi\textsuperscript{3+} to enhance magneto-optical effects \cite{Hansen,Hibiya_1985, zvezdin1997modern}. Previous research on films grown in this way show that the sample is characterized by an uniaxial out-of-plane type anisotropy, as the thin-film shape anisotropy is shown to be overcome by stress-induced anisotropy from a lattice mismatch between substrate and sample \cite{Kubota_2012} together with a small contribution of growth-induced anisotropy due to the site preference of bismuth ions along the growth direction \cite{TmBiFeGaO12_substitutionsgerhard, Growth_anisotropy_euIG}. Consequentially, this gives an ``easy-axis'' along the [$111$] crystallographic direction. The expectations are confirmed by measurements of static magneto-optical Faraday rotation as a function of magnetic field (Supplemental Material \footnote{\label{footnote}See Supplemental Material for magneto-optical characterization of the sample, experimental details of THz generation, pump and probe polarization dependencies, amplitude of dynamics vs THz and external magnetic field, supplemental waveforms and Fourier spectra over a wide temperature range, details on the numerical modelling and comprehensive description of the Lagrangian formalism, which includes Refs. \cite{HoIG_spinflop,PhysRevLett.123.157202,Davydova_2019, Sajadi:15, PhysRev.129.1995}.}).

\begin{figure}[h!]
    \centering
    \includegraphics{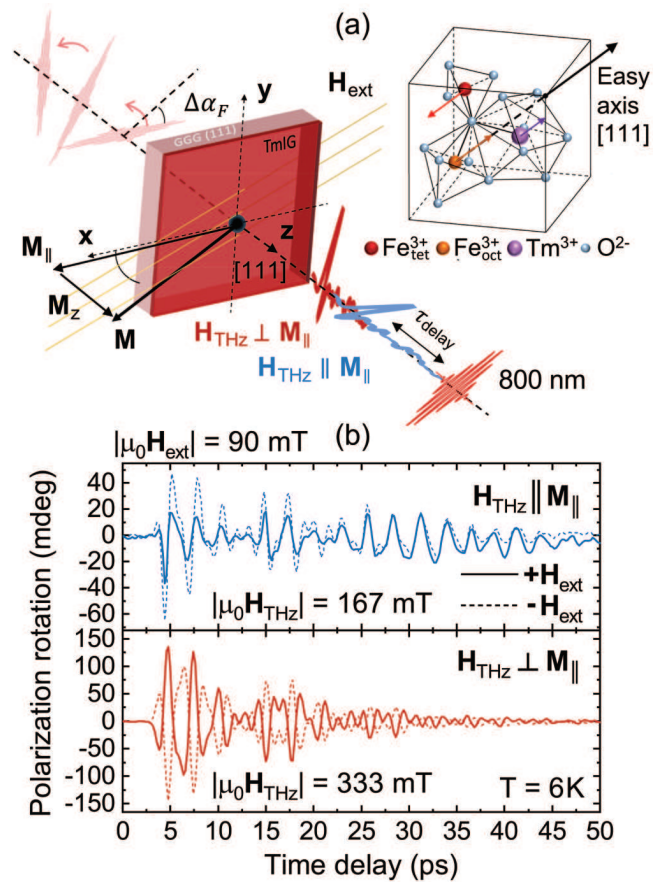}
    \caption{\small{(a) Schematic of the experimental setup. The illustration on the top-right shows the distribution of dodecahedral Tm\textsuperscript{3+} and tetrahedral/octahedral Fe\textsuperscript{3+} ions. Any magnetic moment will tend to align along the [$111$] ``easy-axis''. (b) Polarization rotation $\Delta\alpha_F$ measured as a function of the delay $\tau$ between THz pump and visible probe pulses. Depending on the THz polarization, the mapped dynamics is either odd ($\mathbf{H}_{THz}\parallel y \perp \mathbf{M}_\parallel$) with the external magnetic field or even ($\mathbf{H}_{THz}\parallel x \parallel \mathbf{M}_\parallel $). The measurements were performed at $T=6 $ K.}}
    \label{fig:1}
\end{figure}

In the pump$-$probe experiment, we use optical pulses from a Ti:Sapphire amplifier with a central wavelength of $800$ nm, $4$ mJ energy per pulse, $100$ fs pulse duration and $1$ kHz repetition rate. These pulses were employed to generate single-cycle THz pulses by a titled-front optical rectification technique in a lithium niobate crystal as described in Ref. \cite{Hebling:02} and written in detail in Ref. \cite{Hirori_beamdivergence}. The generated THz beam was tightly focused onto the sample \cite{Hirori_review_4ftheta} and spatially overlapped with a low intensity optical probe beam that was chopped out beforehand from the original beam. Varying time retardation between the THz pump and optical probe pulse, time-resolved measurements were obtained by mapping probe polarization changes induced by the THz pulse using a balanced photo-detector. The strength of the THz electric field was calibrated using the Pockels effect in a thin ($110$)-cut GaP crystal and yields a maximum peak strength of $|\mathbf{E}_{THz}| \approx 1$ MV/cm, implying a peak magnetic field of $0.33$ T. The THz pulse waveform and the corresponding Fourier spectrum are shown in the Supplemental Material. Both the generated THz and optical probe pulses are linearly polarized. The experimental geometry is schematically depicted in Fig. \ref{fig:1}(a). The THz magnetic field is initially along the $x$-axis, but this direction can be controlled by a set of wire-grid polarizers. Note that using this approach, a polarization rotation of $\pi/2$ from the initial state always reduces the THz magnetic field at least by one half. A static external magnetic field $\mu_0\mathbf{H}_{ext}$ of at most $250$ mT was applied at an angle of $\sim 10^\circ$ with the sample plane. Using static Faraday rotation $\alpha_F$ we see that such maximum field strength is sufficient to saturate magnetization in the garnet film.

Figure \ref{fig:1}(b) shows THz-induced ultrafast dynamics of the probe polarization $\Delta \alpha_F$ and how it depends on the THz-pump polarization. By rotating the THz polarization from $\mathbf{H}_{THz}\parallel \mathbf{M}_\parallel$ to $\mathbf{H}_{THz}\perp \mathbf{M}_\parallel$, the symmetry of the high-frequency oscillations with respect to the polarity of the external magnetic field is altered. To reveal the origin of these peculiar THz-induced modulations, we performed systematic studies as a function of pump and probe polarizations, external magnetic field, THz field strength and temperature. 

The observed oscillations of the probe polarization rotation, obviously, are a result of a periodic modulation of optical anisotropy (birefringence) in the sample. A THz pulse is able to induce such optical anisotropy by modifying the dielectric permittivity tensor $\epsilon_{ij}$. If one neglects dissipation, which is a safe approximation for iron garnets at the wavelength of $800$ nm \cite{Wood, zvezdin1997modern}, the tensor is Hermitian \cite{ElectrodynamicsContmedia}. Such type of tensor $\epsilon_{ij}$ can be written as a sum of the symmetric (real) $\epsilon_{ij}^{(s)} = \epsilon_{ji}^{(s)}$ and antisymmetric (imaginary) $\epsilon_{ij}^{(a)}=-\epsilon_{ji}^{(a)}$ parts. Measurements of the THz-induced dynamics as a function of probe polarization angle show no dependency (Supplemental Material \cite{Note1}), indicating that the THz-induced modulations originate from the anti-symmetric part of the dielectric tensor. It means that the polarization rotation $\Delta \alpha_F$ must be assigned to the magneto-optical Faraday effect. In a $[111]$ garnet crystal, this effect is a measure of the magnetization along the $z$-axis \cite{birss1964symmetry,Pisarev_1993}:
\begin{equation}
     \epsilon_{xy}^{(a)} \sim M_z.
    \label{anti}
\end{equation}

When $\mathbf{H}_{THz} \perp \mathbf{M}_\parallel$, changing the external magnetic field polarity from $+\mathbf{H}_{ext}$ to $-\mathbf{H}_{ext}$ flips the sign of the observed dynamics (Fig. \ref{fig:1}(b), red waveforms). Moreover, by increasing the strength of the static magnetic field we found that the amplitude of the oscillations and the net magnetization saturate at the same field (Supplemental Material \cite{Note1}). This fact implies  that the THz-induced dynamics must by assigned to dynamics of the magnetization $\mathbf{M}$. Due to peculiarities of the detection technique (Eq. \eqref{anti}), the measurements are sensitive to modulations of the out-of-plane magnetization. Thus, if we compare the size of the amplitude of the oscillations $\Delta \alpha_F$ with the saturated static Faraday rotation $\alpha_F$ ($\sim 20^\circ$ at $800$ nm), this allows us to quantitatively estimate the relative change of the magnetization along the z-axis during the oscillations $\Delta M_z / M_z \sim 0.012$. For another case $\mathbf{H}_{THz}\parallel \mathbf{M}_\parallel$, the signal also saturates in line with the magnetization $\mathbf{M}$, but the phase of the oscillations is unaffected by the polarity of the external field (see Fig. \ref{fig:1}(b)).
\begin{figure}[h!]
    \centering
    \includegraphics{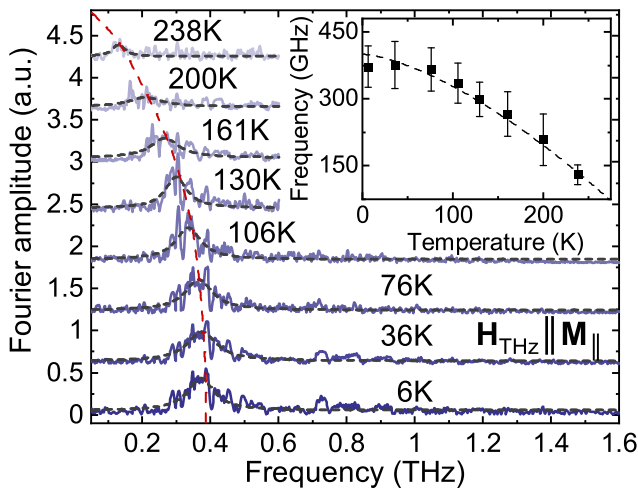}
    \caption{\small{Fourier spectrum of the THz-induced signal ($\mathbf{H}_{THz}\parallel \mathbf{M}_\parallel$) measured at various temperatures. Central frequencies of the peaks deduced from the fit are plotted as a function of temperature in the inset. The dotted line denotes a fit with Eq. \eqref{fitfreq} ($\omega_0 = 400 $ GHz, $T_C = 314 $ K) and the bars denote $\pm$ half-width-half-maximum of the fitted Lorentzians. The FFT spectrum for $\mathbf{H}_{THz}\perp \mathbf{M}$ is added to the Supplemental Material \cite{Note1}.}}
    \label{fig:fig2}
\end{figure}

\begin{figure}
    \centering
    \includegraphics[width = \linewidth]{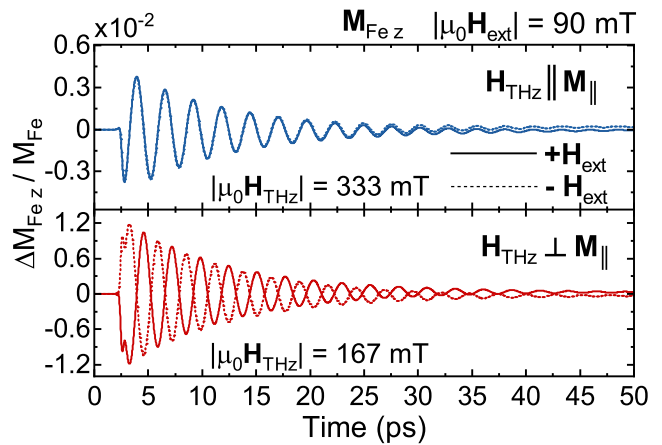}
    \caption{\small{Dynamics in the $z$-component of iron $\mathbf{M}_{Fe}$ modeled by LLG equations.}}
    \label{fig:fig4}
\end{figure}

Figure \ref{fig:fig2} shows the Fourier spectra of the THz-induced waveforms ranging in the entire accessible temperature range when $\mathbf{H}_{THz}\parallel \mathbf{M}_\parallel $. The inset summarizes the temperature dependence of the peak frequency, and this behaviour is in qualitative agreement with what could be expected for an exchange mode in rare-earth iron garnets \cite{PhysRev.129.1995, PhysRevLett.105.107402}. In order to get a better insight into the THz-induced magnetization dynamics, we modelled the response with the help of the Landau-Lifshitz-Gilbert (LLG) equations \cite{Kirilyuk_rev}. The equations, in particular, account for rFDT derived by \cite{Mondal2016}:
\begin{equation}
    \frac{d\mathbf{M_i}}{dt} = - \gamma_i \mathbf{M_i} \times \mathbf{B}_{i}^{eff}(t) + \frac{\alpha_i}{M_i}\mathbf{M_i} \times \Big(\frac{d\mathbf{M_i}}{dt} + \frac{a_i^3}{\mu_B}\frac{d \mathbf{H}}{d t}\Big),
    \label{LLG}
\end{equation}
where $i =$ Fe, Tm. We use literature $g$-values for thulium $g_{Tm} = 7/6$ and iron $g_{Fe} =  2$ \cite{wohlfarth1986handbook}. 
Based on the Ga-content, the sublattice magnetization of iron $|\mathbf{M}_{Fe}| = 4.2$ ($\mu_B$ per formula unit R\textsubscript{3}Fe\textsubscript{5}O\textsubscript{12}) \cite{wohlfarth1986handbook} is antiferromagnetically coupled to the magnetization of thulium $|\mathbf{M}_{Tm}| = 2 $. The latter is taken to match the effective $g$-factor $g_{ef} \equiv (M_{Fe}-M_{Tm})/((M_{Fe}/g_{Fe}) - M_{Tm}/g_{Tm}) \approx 6$ measured in this sample (Supplemental Material \cite{Note1}). The volume of the unit cell $a_i^3$ \footnote{We have used the following set of values for calculating the magnitude of rFDT terms: $a^3_{Fe} = 1.221 \times 10^{-28}\text{m}^3$, $a^3_{Tm} = 5.815 \times 10^{-29}\text{m}^3$ and vacuum permeability $\mu_0 = 1.257 \times 10^{6}\text{T m A}^{-1}$. } per spin constitutes a small factor $a^3/\mu_B \sim 10^{-5}$ m/A. The effective magnetic fields $\mathbf{B}_i^{eff} \equiv - \delta \Phi/ \delta \mathbf{M}_i $ (in T) are derived from the thermodynamic potential $\Phi$ \cite{Kirilyuk_rev}, containing exchange interaction and Zeeman coupling to the external field and THz magnetic field $\mathbf{H}(t)$ (in A/m). For the model we use a realistic exchange constant $\Lambda = -30$ T/$\mu_B$ \cite{TmIG_molecularfield, Molecularfields, wohlfarth1986handbook} and THz magnetic field modelled by the Gaussian derivative function fitted to the experimental waveform (see Supplemental Material \cite{Note1}). The initial state of the net magnetization vector is taken along the external field, considering we saturated the magnetization experimentally. The numerical solution of these equations reveals that the THz magnetic field induces dynamics of the Néel vector $\mathbf{L} \equiv \mathbf{M}_{Fe} - \mathbf{M}_{Tm} $ and the magnetization $\mathbf{M} \equiv \mathbf{M}_{Fe} + \mathbf{M}_{Tm}$. The dynamics of $\mathbf{M}_{Fe}$, which dominates the detected magneto-optical signal, is shown in Fig. \ref{fig:fig4}. The phenomenological Gilbert damping factors of $\alpha_{Fe}/M_{Fe} = \alpha_{Tm}/M_{Tm} = 0.0015$ have been taken to match the experimental observations.
The simulation contains a high-frequency magnetic resonance at around $380$ GHz, which we identify as the Kaplan-Kittel exchange mode since its frequency depends linearly on the exchange constant \cite{doi:10.1063/1.1699018}. The dynamics of $M_{Fe,z}(t)$ in Fig. \ref{fig:fig4} is in agreement with our experimental results in Fig. \ref{fig:1}(b). It has a larger amplitude and changes sign upon reversing $\mathbf{M}$ when $\mathbf{H}_{THz} \perp \mathbf{M}_\parallel$, while the sign is conserved if $\mathbf{H}_{THz} \parallel \mathbf{M}_\parallel$. The amplitude matches very well to the experimental values even if the rFDT term is not taken into account. As proposed in Ref. \cite{Mondal2019} the contribution of this term will be indeed small in cases of low damping $\alpha_{1,2} < 0.01$. Altogether, the simulations point out that the observed oscillations correspond to the exchange mode of spin resonance and show that Zeeman-torque plays the dominant role in the excitation of this mode with THz magnetic field.

These conclusions can also be confirmed analytically using Lagrangian mechanics and the effective Lagrangian (see Supplemental Material \cite{Note1} for full derivation):
\begin{eqnarray}
\begin{aligned}
\label{Leff}
\mathcal{L}_{eff} &=  \frac{\mathcal{M}^2}{2\delta}\Bigg[\left(\Big(\frac{\dot\phi}{\overline{\gamma}} - H\Big)\sin\theta + h_y\cos\theta\cos\phi \right)^2 \\&\quad + \left(\frac{\dot\theta}{\overline{\gamma}} + h_y\sin\phi\right)^2 \Bigg] 
+  m \Big(H - \frac{\dot\phi}{\gamma_{ef}}\Big) \cos\theta \\&\quad + mh_y\sin\theta\cos\phi + K_U\sin^2\theta \sin^2\phi.
\end{aligned}
\end{eqnarray} \newline
Where $\mathcal{M} = M_{Fe} + M_{Tm}$, $m = M_{Fe} - M_{Tm}$, $\delta \equiv -4\Lambda M_{Fe} M_{Tm}$, $1/\overline{\gamma} = (M_{Fe}/\gamma_{Fe} + M_{Tm}/\gamma_{Tm})/(M_{Fe} + M_{Tm})$, $\gamma_{ef} = g_{ef} \mu_B/\hbar$,  $h_x(t)$ and $h_y(t)$ are the THz magnetic field components in the sample $x-y$ plane as in Fig. \ref{fig:1} and $H(t) \equiv H_{ext} + h_x(t)$ the total field along the external field $x$-direction (here $10^{\circ}$ inclination angle of external magnetic field is ignored). The polar angle $\theta \in [0,\pi]$ is defined with respect to the external field $x$-axis. 
In this coordinate system, the net magnetization vector can be expressed as $\mathbf{M} = m(\cos\theta, \sin\theta \cos\phi, \sin\theta\sin\phi)$. Equations of motion now follow from Euler-Lagrange equations, taking into account a phenomenological damping term through a Rayleigh function \cite{Davydova_2019}. The results can be linearized about the ground state angles $\theta_0, \phi_0$, found by minimization of the thermodynamic potential $\Phi$ for which we find $\phi_0 = \pi/2$ and $\theta_0$ depending on the ratio of external field to anisotropy. This has been done for general $\theta_0$ in the Supplemental Material \cite{Note1}, yielding complex equations of motion. In the special case of zero external field, the spins lie along the easy axis of anisotropy ($\theta_0 = \pi/2$). Linearizing around the ground-state angles $\theta = \theta_0 + \theta_l$, $\phi = \phi_0 + \phi_l$ with $\theta_l, \phi_l \ll 1$, the equations of motion then take the simple form:
\begin{multline}
    \label{motiontheta}
    \ddot{\theta}_l + \frac{\alpha\mathcal{M}\overline{\gamma}}{\chi_\perp}\dot{\theta}_l +  \frac{2K_U\overline{\gamma}^2}{\chi_\perp}\theta_l - \frac{m\overline{\gamma}^2}{\gamma_{ef}\chi_\perp}\dot\phi_l  = - \overline{\gamma}\dot h_y - \frac{m\overline{\gamma}^2h_x}{\chi_\perp},
\end{multline}
\begin{multline}
\label{motionphi}
    \ddot\phi_l + \frac{\alpha \mathcal{M}\overline{\gamma}}{\chi_\perp}\dot\phi_l +\frac{2K_U\overline{\gamma}^2}{\chi_\perp}\phi_l  + \frac{m\overline{\gamma}^2}{\gamma_{ef}\chi_\perp}\dot\theta_l = \overline{\gamma} \dot h_x  - \frac{m\overline{\gamma}^2h_y}{\chi_\perp}.
\end{multline}
Here $\chi_\perp \equiv \frac{\mathcal{M}^2}{\delta}$ is a constant inversely proportional to the exchange constant. It is seen that the large THz field derivative term $\gamma \dot{h}_i$ appears as the dominant driving force, in accordance with our understanding how dynamical THz fields may excite antiferromagnetic magnons in antiferromagnets (where $m \to 0$) by Zeeman interaction \cite{PhysRevLett.123.157202, KIMEL20201}. Moreover, each equation of motion contains a mutually orthogonal component of the field-derivative $\dot{h}_{x,y}$. Noting that $\mathbf{H}_{THz} \perp \mathbf{M}_\parallel$ leads to $\dot h_x = 0$ and $\mathbf{H}_{THz} \parallel \mathbf{M}_{\parallel}$ to $\dot h_y = 0$, the symmetry with respect to external field $\pm\mathbf{H}_{ext}$, as observed experimentally, can now be explained (see Supplemental Material \cite{Note1}).

Moreover, considering free precession $\alpha \to 0$, $h_{x,y} \to 0$, the absolute eigenfrequencies of the coupled set of equations \eqref{motiontheta} - \eqref{motionphi} are:
\begin{eqnarray}
\label{KK}
    \omega_{ex} &=& \frac{m\overline{\gamma}^2}{\gamma_{ef} \chi_\perp} \approx |\Lambda|(|\gamma_{Tm}|M_{Fe} - |\gamma_{Fe}|M_{Tm}),\\
    \label{FM}
    \omega_{FM} &=& \gamma_{ef}\frac{2K_U}{m} \equiv \gamma_{ef} H_{a}.
\end{eqnarray}
Equation \eqref{KK} corresponds to Kaplan-Kittel's exchange resonance frequency \cite{doi:10.1063/1.1699018} while Eq. \eqref{FM} describes the conventional ferromagnetic precession of the net magnetization in the anisotropy field $H_a$. Using Eq. \eqref{KK} and Bloch's law for the spontaneous magnetization of iron while $M_{Tm}(T) \sim M_{Fe}(T)$, we fitted the temperature dependence of the oscillations frequency shown in inset of Fig. \ref{fig:fig2} using:
\begin{equation}
    \label{fitfreq}
    \omega_{ex}(T) \sim \omega_0\left(1-\left(T/T_C\right)^{\frac{3}{2}}\right).
\end{equation}
where $\omega_0$ the exchange resonance frequency at zero Kelvin. In reality $M_{Tm}$ drops faster with temperature than the magnetization of iron, accounting for the slight rise of frequency at low temperatures. In general, the fit is another confirmation of the validity of our assumption that the observed oscillations correspond to the exchange mode.

In conclusion, investigating the response of ferrimagnets to THz fields and comparing the data with theoretical predictions from numerical solutions of the Landau-Lifshitz-Gilbert equations and analytical solutions derived from Euler-Lagrange equations of motion, we showed that the THz field excites the exchange mode in the ensemble of antiferromagetically coupled spins. We demonstrated that the Zeeman-torque plays a dominant role in the coupling of the THz-field to the spins. While quantitative studies of spin dynamics in compensated antiferromagnets seem to require complex magnetometry techniques, ferrimagnets facilitate an excellent playground to study dynamics of antiferromagnetically coupled spins. At last, we would like to point out that previous measurements of ferrimagnetic resonance \cite{GdFeCo2017, PhysRevLett.105.107402} could only reveal an effective gyromagnetic ratio. Using excitation of exchange mode with THz magnetic field, magneto-optical detection via the Faraday effect and comparison of the observed amplitudes of magnetization dynamics with the results of numerical simulations provides a universal technique to directly estimate the individual gyromagnetic ratio of the ions. 
\begin{acknowledgments}
The authors thank S. Semin, Ch. Berkhout and P. Albers for technical support. The work was supported by de Nederlandse Organisatie voor Wetenschappelijk Onderzoek (NWO). M.V.L. acknowledges the support from the Russian Foundation for Basic Research (Nos. 18-29-27020 and 18-52-16006).
\end{acknowledgments}



\nocite{*}

\bibliography{ms}

\end{document}


\title{\centering\normalsize\bfseries\MakeUppercase{Supplemental Material to: THz Field-induced Spin Dynamics in Ferrimagnetic Iron Garnets}}
 \date{\vspace{-13ex}}
 \maketitle
\section{Static magneto-optical characterization of TmIG sample}
\begin{figure}[h!]
    \centering
    \vspace{-4ex}
    \includegraphics[width = 0.75\textwidth]{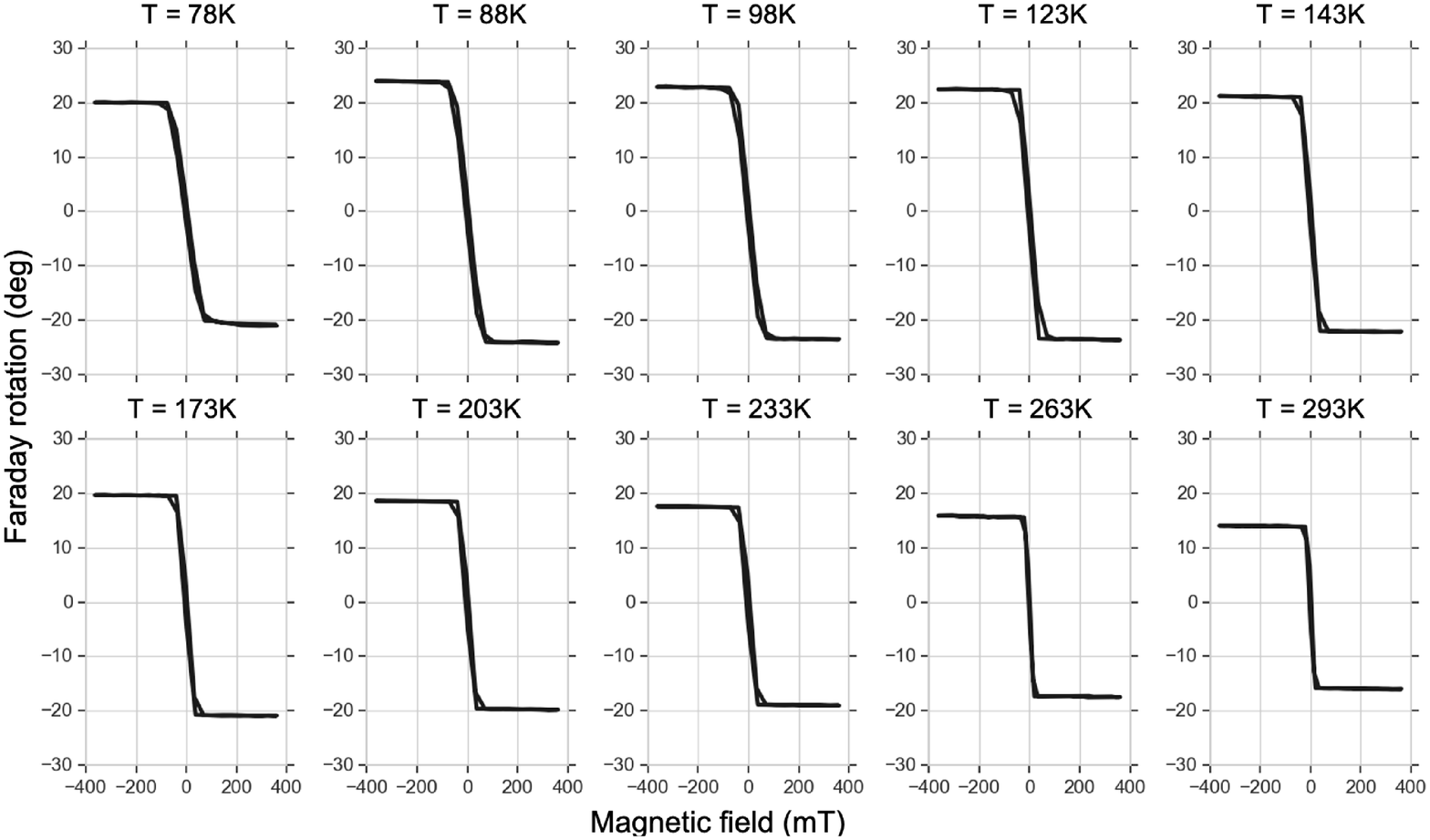}
    \vspace{-1ex}
    \caption{\small{Measurements of the magneto-optical Faraday rotation using a continuous-wave helium-neon laser ($\lambda = 632.8$ nm) with both the external field and the light's wave-vector perpendicular to the sample plane. A paramagnetic contribution $\sim 0.56$ deg/T from the cryostat glass windows has been subtracted from the raw data. The data exhibits a large rotation and demonstrates a weak easy-axis type of anisotropy with small coercive $< 6$ mT and saturation $< 25$ mT field. No compensation point was observed above nitrogen temperatures $> 77$ K.}}
    \label{fig:staticsHeNe}
\end{figure}

\begin{figure}[h!]
\vspace{-2ex}
    \centering
    \includegraphics[width = 0.6\textwidth]{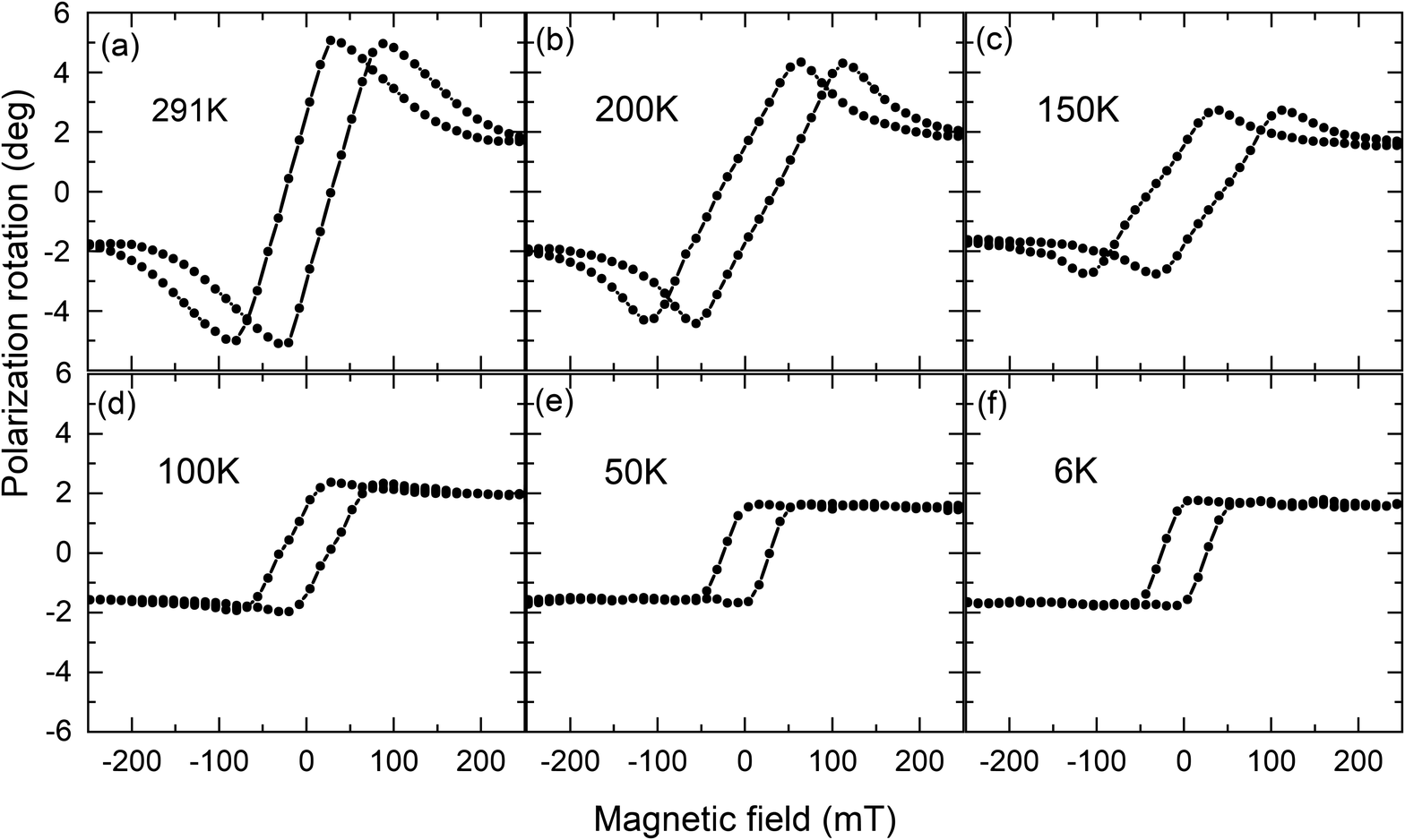}
    \caption{\small{Static polarization rotation measurements with light at the wavelength of $800$ nm in the experimental geometry (see Fig. 1(a) in article). The evolution of hysteresis loop form can be due to temperature dependent anisotropy constants. Clearly, no magnetization compensation point is observed in this temperature range. At T $ = 6$ K, the saturated polarization rotation is $
    \sim \pm 1.65^\circ$, and given the angle of the magnetic field of $10^\circ$, this has been used to estimate the Faraday rotation ($\sim \pm 10^\circ)$ when the magnetization is along the sample normal. }}
    \label{fig:supfig2}
\end{figure}
\ \ \ \ \ \
\newpage

\begin{figure}[h!]
    \centering
    \includegraphics{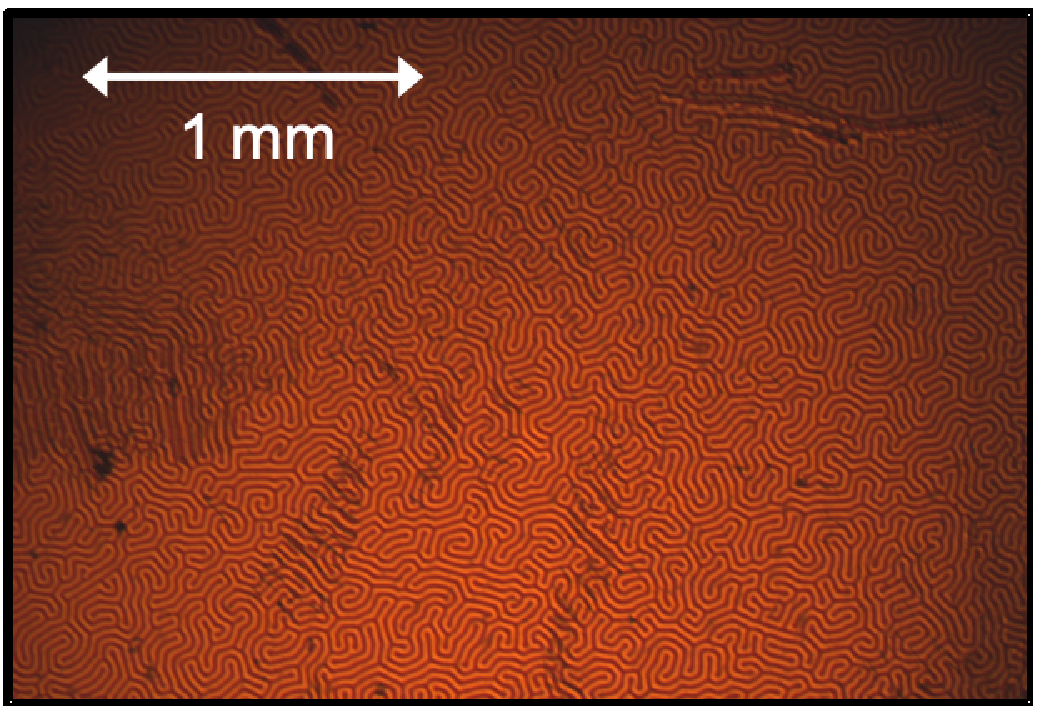}
    \caption{\small{Domain pattern seen by magneto-optical microscopy in transmission at zero field. The typical "labyrinth" type domains grow with decreasing temperature, which indicates a growing role of easy axis anisotropy and thulium magnetization \cite{HoIG_spinflop}. When applying an external magnetic field along the out-of-plane easy axis, the domains along this field expand and the sample will be uniformly magnetized for relatively small field (see Suppl. Fig. \ref{fig:staticsHeNe})}}
    \label{fig:domains}
\end{figure}
\ \ \ \ 
\newpage

\section{Experimental setup}
The experimental setup regarding THz generation by optical rectification in lithium niobate is described in detail in \cite{PhysRevLett.123.157202,Sajadi:15}. The THz path was purged with nitrogen to avoid water absorption lines in the THz spectrum.  A small part of the initial $800$ nm beam is chopped out beforehand (ratio $1:100$) and is brought to spectral and temporal overlap with the THz pump pulse. The focused spot size of the probe beam is considerably smaller than that of the THz. The waveform of the THz pulse was mapped using electro-optical sampling in a $50$ $\mu$m GaP [$110$] crystal seen in Supplemental Fig. \ref{fig:THz1}.  
\begin{figure}[h!]
    \centering
    \includegraphics{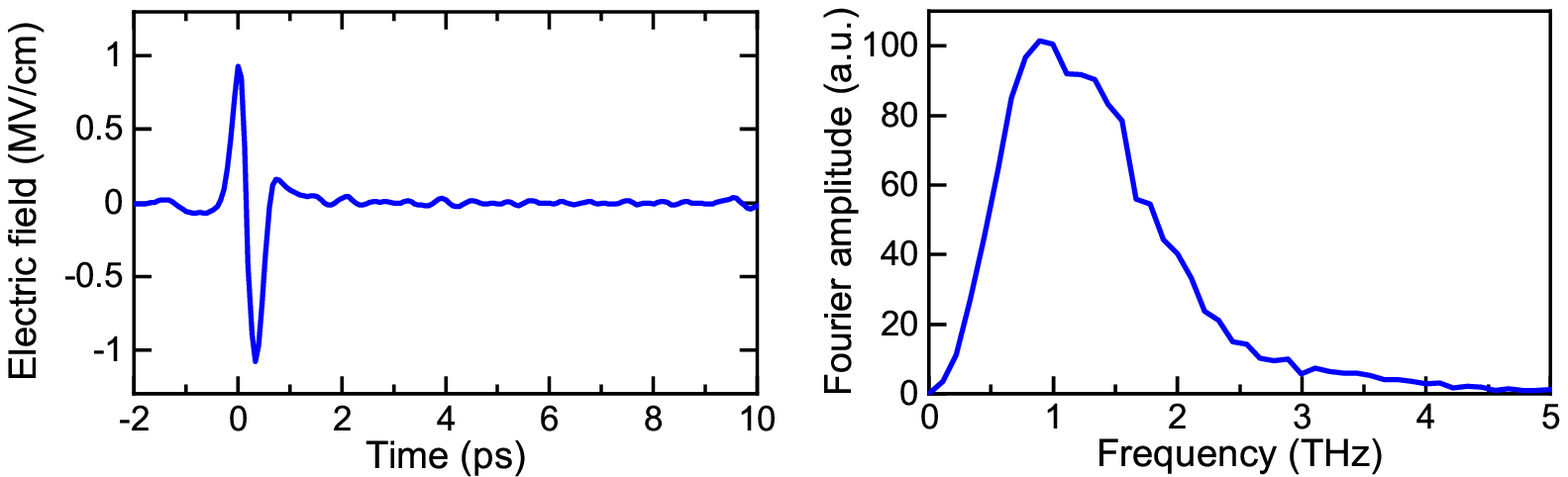}
    \caption{\small{THz waveform and corresponding Fourier spectrum measured by EO sampling in GaP. }}    \label{fig:THz1}
\end{figure}
\section{Supplemental Results}
\begin{figure}[h!]
    \centering
    \includegraphics{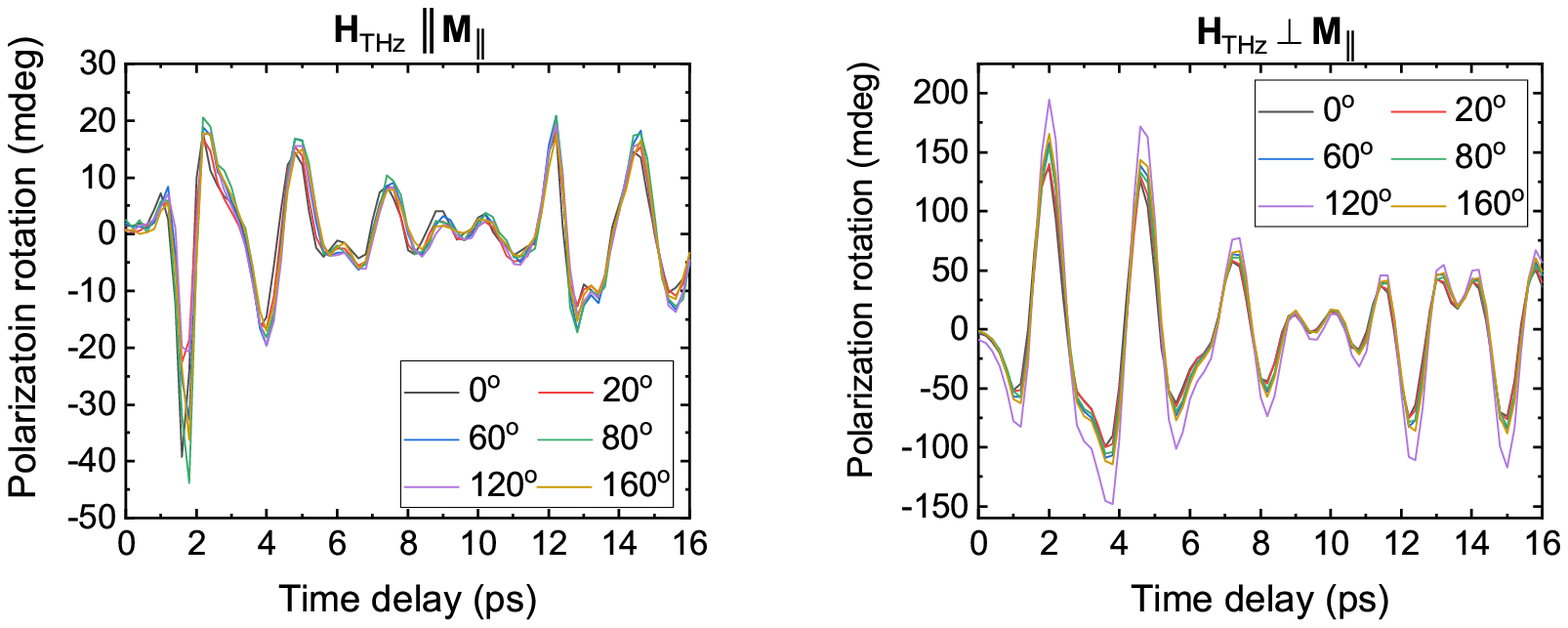}
    \caption{\small{THz induced polarization rotation waveforms for two orthogonal THz pump polarizations (two figures) and for several orientations of the probe polarization. The angle depicted is the angle of the probe electric field with respect to the experimental $x$-axis (Fig. 1(a) of the article). This data implies that the THz induced signals are Faraday rotation (see main text).  }}
    \label{fig:supfig5}
\end{figure}

\begin{figure}
    \centering
    \includegraphics{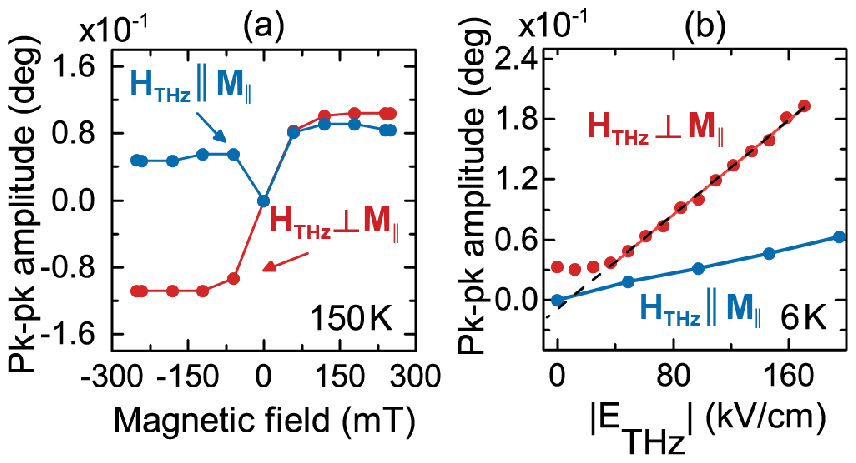}
    \caption{\small{Peak-to-peak amplitudes of THz induced waveforms as a function of external magnetic field (a) and THz field (b) for two orthogonal THz pump polarizations. Bending of the red dots at low THz fields is attributed to the facts that the THz light is not perfectly linearly polarized and imperfections of the wire grid polarizers.} }
    \label{fig:supfig6}
\end{figure}
\newpage
\begin{figure}[h!]
    \centering
    \includegraphics{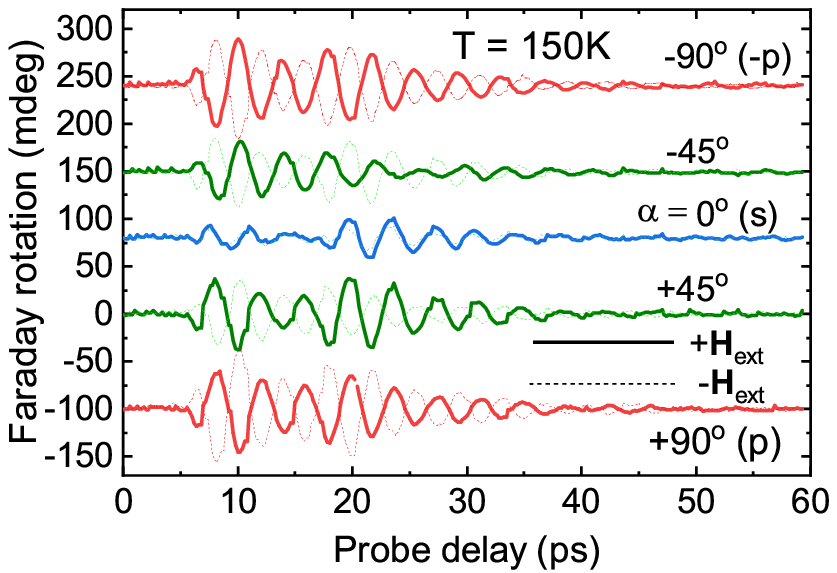}
    \caption{\small{THz induced Faraday rotation waveforms obtained at several angles of the pump polarization angle $\alpha$ and applied external field of $\pm 250$ mT. Besides the previous figure, this is the only graph where a weaker THz electric field of $160$ kV/cm has been used. The result shows that measuring at $\alpha = \pm45^\circ$, in between the fully symmetric $\alpha=0^\circ$ and fully antisymmetric orthogonal $\alpha = \pm90^\circ$, results in a mix of symmetry/antisymmetry. Moreover, the effects gradually become weaker towards $\alpha = 0^\circ$ ($\mathbf{H}_{THz} \parallel \mathbf{M}_\parallel$).}}
    \label{fig:supfig7}
\end{figure}

\newpage

\begin{figure}[h!]
    \centering
    \includegraphics{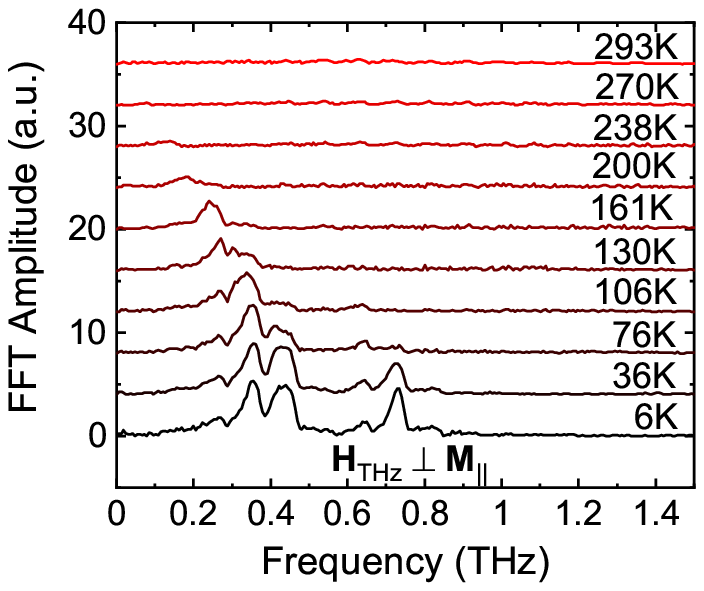}
    \caption{\small{FFT spectra of THz induced Faraday rotation for $\mathbf{H}_{THz} \perp \mathbf{M}_{\parallel}$. The exchange-mode frequency at about $375$ GHz shows softening similar to the case $\mathbf{H}_{THz} \parallel \mathbf{M}_\parallel$ as is presented in article Fig. 2. At lower temperatures another high frequency ($725$ GHz) appears. It is known that crystal field transition may appear in this region \cite{PhysRev.129.1995}, but if it can be attributed to these transitions is yet unclear. }}
    \label{fig:supfig8}
\end{figure}

\begin{figure}[h!]
    \centering
    \includegraphics{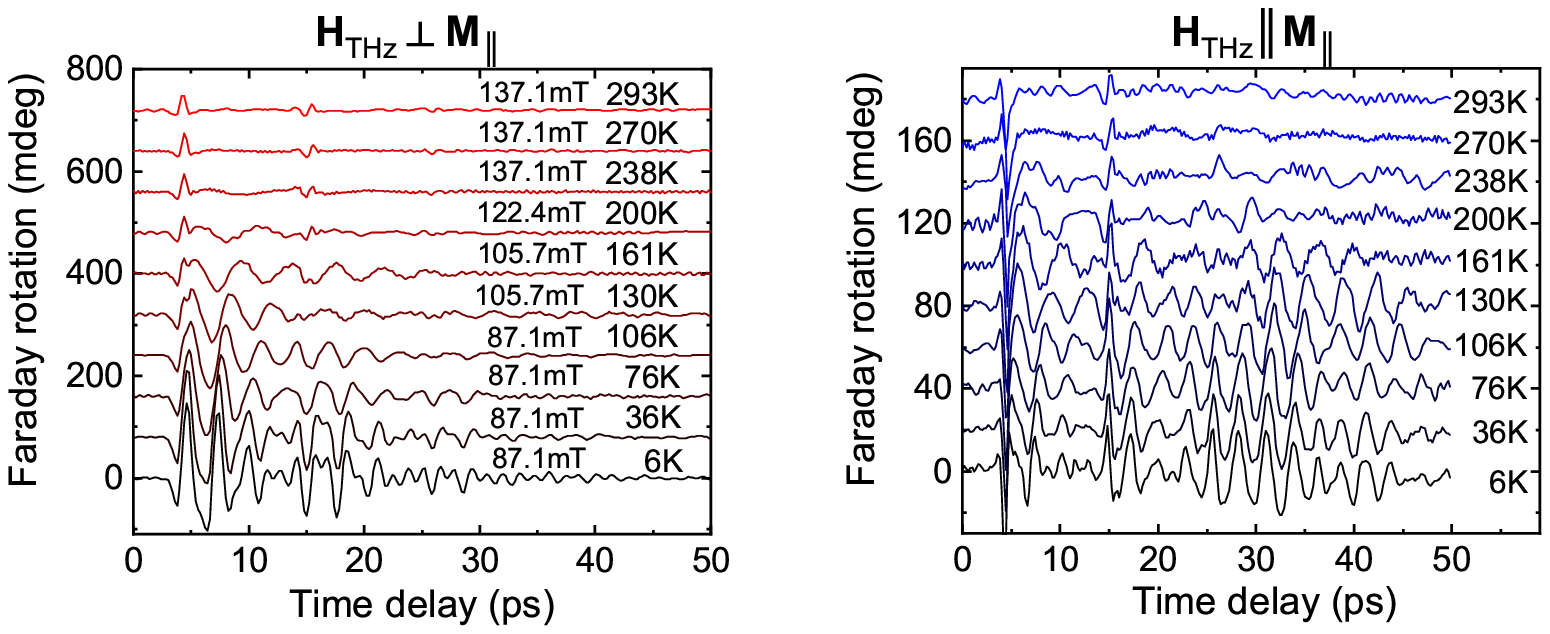}
    \caption{\small{Experimental waveforms of THz induced Faraday rotation as a function of temperature. In both cases the same external fields (specified in the first figure) have been applied to ensure saturation of static magnetization.}}
    \label{fig:supfig9}
\end{figure}
 \newpage
\begin{figure}
    \centering
    \includegraphics{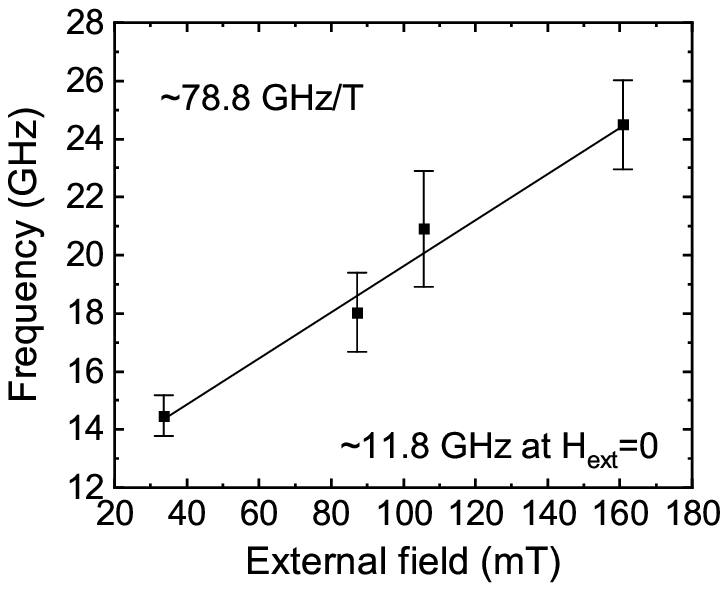}
    \caption{\small{Preliminary data of THz-induced ferromagnetic resonance, used to estimate the effective $g$-factor $g_{eff} \approx 6$.}}
    \label{fig:geff}
\end{figure}
\begin{figure}[h!]
    \centering
    \includegraphics{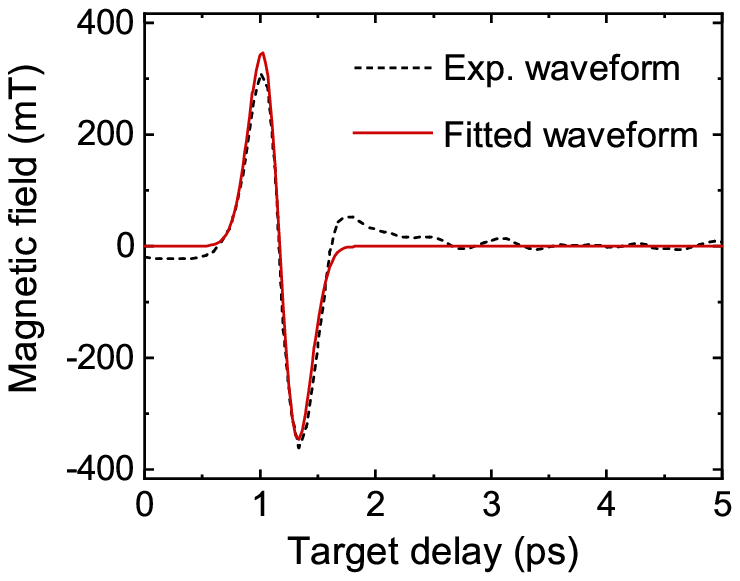}
    \caption{\small{The dotted line shows the experimentally calibrated THz magnetic field pulse, which has been fitted using the Gaussian derivative function $G'(x) = -2A((x-d)/w) \exp\left[((x-d)/w)^2\right]$ with $A = 404$ mT, $d = 1.17$ ps (variable, determines arrival time of pulse) and $w = 0.2223$ ps pulse-width.}}
    \label{fig:THz}
\end{figure}
\newpage
\begin{figure}
    \centering
    \includegraphics{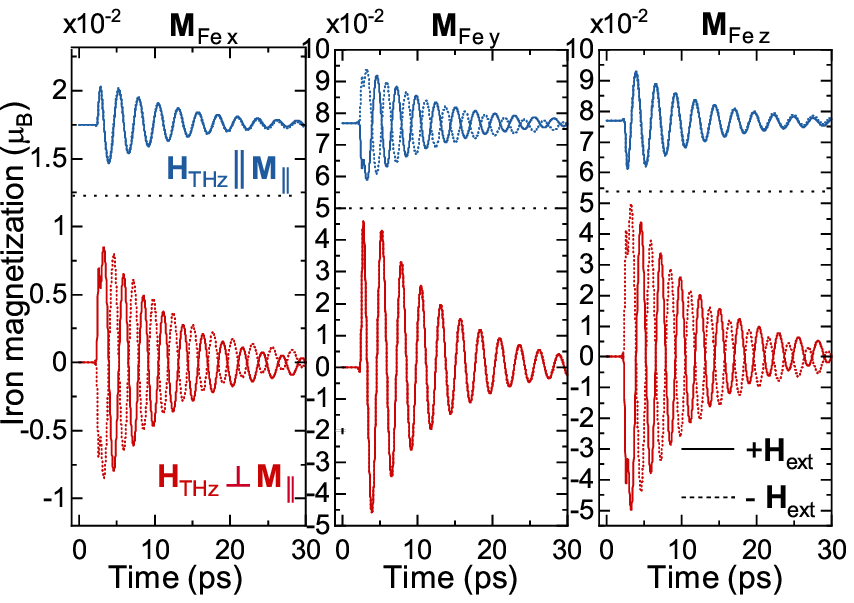}
    \caption{\small{Simulated dynamics of the iron magnetization $\mathbf{M}_{Fe}$ using LLG equations and plotted separately for the $x$, $y$ and $z$ components where $z$ coincides with the sample
out-of-plane axis. It shows how the symmetry with respect to external field is exactly opposite when looking at the $y$-component, to which we are not experimentally sensitive.}}
    \label{fig:supfig12}
\end{figure}
\section{Equations of motion derived from Lagrangian formalism}
\begin{figure}[h!]
    \centering
    \includegraphics[scale = 1]{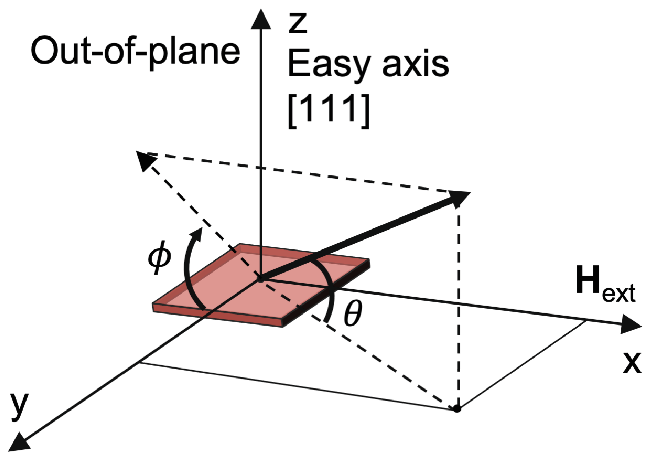}
    \caption{\small{Coordinate system used for Lagrangian equation.}}
    \label{fig:coords}
\end{figure}

We start from the following Lagrangian and Rayleigh dissipation functions, which are equivalent to the LLG equations for a two-sublattice ferrimagnet \cite{Davydova_2019}:
\begin{eqnarray}
    \label{Lagrangian}
    \mathcal{L} &=& T - \Phi \nonumber\\
    &= &-\frac{M_{Fe}}{\gamma_{Fe}}\cos\theta_{Fe}\frac{\partial\phi_{Fe}}{\partial t} -\frac{M_R}{\gamma_{R}}\cos\theta_R\frac{\partial\phi_R}{\partial t} - \Phi \\
    \mathcal{R} &=& \mathcal{R}_{Fe} + \mathcal{R}_R, \text{   \ \ \ \ \ \ \ \   } \mathcal{R}_{Fe,R} = \frac{\alpha M_{Fe,R}}{2\gamma_{Fe,R}}\big(\dot{\theta}^2_{Fe,R} + \sin^2\theta_{Fe,R}\dot{\phi}^2_{Fe,R} \big),
\end{eqnarray}
where $\theta_i$ and $\phi_i$ the polar and azimuthal angles of the iron (Fe) and rare-earth (R) sublattices in the experimental coordinate system with the $x$-axis aligned to the external magnetic field $\mathbf{H}_{ext}$ (see Fig. \ref{fig:coords}, here we ignore the $10^\circ$ inclination of the field for simplification). 

The thermodynamic potential used is:
\begin{eqnarray}
    \label{phi}
    \Phi &=& -(\mathbf{M}_{Fe} + \mathbf{M}_{R}) \cdot\mathbf{H}_{ef} - \Lambda \mathbf{M}_{Fe}\cdot\mathbf{M}_{R} - K_{Fe}\frac{(\mathbf{M}_{Fe}\cdot\mathbf{n})^2}{M_{Fe}^2} - K_{R}\frac{(\mathbf{M}_R\cdot\mathbf{n})^2}{M_R^2}.
\end{eqnarray}
Here $\mathbf{n} = (0,0,1)$ is the directional vector of the easy axis of anisotropy, $\Lambda < 0$ the intersublattice exchange constant and $K_{Fe,R}>0$ the uniaxial anisotropy constants. The Euler-Lagrange equations w.r.t. $\theta_i$, $\phi_i$ give rise to four (coupled) equations of motion (two for each sublattice), and this is generally not easy and sometimes even impossible to solve. Instead, in Ref. \cite{Davydova_2019} an effective Lagrangian is obtained by assuming the canting of the two sublattices are equal and are assumed to be small. This approach generally works at field well below the exchange field (small canting), and it is valid here as the static measurements indicate we are well below the spin-flop field.

We introduce the usual definitions of the magnetization $\mathbf{M} = \mathbf{M}_{Fe} + \mathbf{M}_R$  and antiferromagnetic (Néel) vector $\mathbf{L} = \mathbf{M}_{Fe} - \mathbf{M}_R$. These two vectors are parameterized using a set of angles $\theta,\epsilon$  and $\phi, \beta$  defined as:
\begin{eqnarray}
\theta_{Fe} &=& \theta - \epsilon, \quad \theta_{R} = \pi-\theta-\epsilon, \label{theta} \\
\phi_{Fe} &=& \phi + \beta, \quad \phi_{R} = \pi + \phi -\beta. \label{phi_angle}
\end{eqnarray}
In the quasi-antiferromagnetic approximation \cite{Davydova_2019}, the canting angles are assumed to be small $\epsilon \ll 1$, $\beta \ll 1$. In first order approximation, the $\mathbf{M}$ and $\mathbf{L}$ are then naturally defined as: 
\begin{eqnarray}
\mathbf{M} &=& m(\cos\theta, \sin\theta\cos\phi, \sin\theta\sin\phi) \\
\mathbf{L} &=& \mathcal{M}(\cos\theta, \sin\theta\cos\phi, \sin\theta\sin\phi)
\end{eqnarray}
where $m \equiv M_{Fe} - M_{R}$ and $\mathcal{M} \equiv  M_{Fe} + M_{R} $.
Substituting our new set of angles (\ref{theta})-(\ref{phi_angle}) into the Lagrangian (\ref{Lagrangian}) and expanding up to quadratic terms in the small variables $\epsilon, \beta$ gives for the kinetic energy part:
\begin{eqnarray}
\mathcal{L} &=&  -\frac{m}{\gamma_{ef}}\dot{\phi}\cos\theta - \frac{\mathcal{M}}{\overline{\gamma}}\sin\theta \left(\dot{\phi}\epsilon + \beta \dot{\theta}\right) - \Phi \label{Lexpanded}
\end{eqnarray}
where we defined:
\begin{equation}
    \frac{1}{\gamma_{ef}} \equiv \frac{M_{Fe}/\gamma_{Fe} - M_{R} / \gamma_R}{M_{Fe} - M_R} \ \ \ \ \ \ \ \ \ \text{     and      } \ \ \ \ \ \ \ \ \ \ \frac{1}{\overline{\gamma}} \equiv \frac{M_{Fe}/\gamma_{Fe} + M_{R} / \gamma_R}{M_{Fe} + M_R}.
\end{equation}
The potential energy $\Phi$ can be expanded similarly. Here, we make the simplification that both sublattices experience the same effective anisotropy $K_U \equiv (K_{Fe} + K_R)/2$ in which case the anisotropy terms can by replaced by a single term $-K_U(\mathbf{l}\cdot\mathbf{n})^2$ where $\mathbf{l} = \mathbf{L}/|\mathbf{L}|$. Furthermore, the effective field $\mathbf{H}_{ef}$ in (\ref{phi}) consists of the static external field and the time-dependent THz magnetic field $\mathbf{H}_{ef} = \mathbf{H}_{ext} + \mathbf{H}_{THz}$. The external field is chosen along the $x$-axis $\mathbf{H}_{ext} = (H_0,0,0)$, while we assume the THz magnetic field lies in the $x-y$ plane $\mathbf{H}_{THz} \equiv (h_x(t), h_y(t), 0$) (see Fig. 1 from the article). Writing $\delta \equiv -4\Lambda M_{Fe} M_{R}$ and $H \equiv H_{ext} + h_x$, the potential energy becomes after expanding in $
\epsilon, \beta$:
\begin{eqnarray}
    \label{Phiexpanded}
\Phi &=& -mH\cos\theta - \mathcal{M}H\epsilon\sin\theta - mh_y \sin\theta\cos\phi + \mathcal{M}h_y\beta \sin\theta\sin\phi \\ 
 &+& \mathcal{M}h_y \epsilon \cos\theta \cos\phi - mh_y\cos\theta\sin\phi \ \epsilon\cdot\beta +\frac{\delta}{2}\left(\epsilon^2 + \beta^2 \sin^2\theta \right) - K_U\sin^2\theta \sin^2\phi.
\nonumber \end{eqnarray}
We will ignore the term containing $\epsilon \cdot \beta$ as it is very small, from the quadratic terms only the ones proportional to the exchange constant $\sim \delta$ survive. We exclude the variables $\epsilon$, 
$\beta$ by solving the Euler-Lagrange equations $\frac{d}{dt}\frac{\partial \mathcal{L}}{\partial \dot{\epsilon}} - \frac{\partial \mathcal{L}}{\partial \epsilon} = -\frac{\partial \mathcal{R}}{\partial\dot{\epsilon}} \approx 0$ and $\frac{d}{dt}\frac{\partial \mathcal{L}}{\partial \dot{\beta}} - \frac{\partial \mathcal{L}}{\partial \beta} = -\frac{\partial \mathcal{R}}{\partial\dot{\beta}} \approx 0$, giving:
\begin{eqnarray}
\epsilon &=& \frac{\mathcal{M}}{\delta } \sin\theta \left(H - \frac{\dot\phi}{\overline{\gamma}}\right) - \frac{\mathcal{M}h_y}{\delta}\cos\theta\cos\phi, \\
\beta \sin\theta &=& -\frac{\mathcal{M}}{\delta}\left(\frac{\dot\theta}{\overline{\gamma}} + h_y \sin\phi\right).
\end{eqnarray}
Substituting in \eqref{Lexpanded}-\eqref{Phiexpanded} and rearranging terms yields the effective Lagrangian from the article:
\begin{eqnarray}
\begin{aligned}
\label{Leff}
\mathcal{L}_{eff} &=  \frac{\mathcal{M}^2}{2\delta}\Bigg[\left(\Big(\frac{\dot\phi}{\overline{\gamma}} - H\Big)\sin\theta + h_y\cos\theta\cos\phi \right)^2 + \left(\frac{\dot\theta}{\overline{\gamma}} 
+ h_y\sin\phi\right)^2 \Bigg] 
+  m \Big(H - \frac{\dot\phi}{\gamma_{ef}}\Big) \cos\theta \\&\quad + mh_y\sin\theta\cos\phi + K_U\sin^2\theta \sin^2\phi.
\end{aligned}
\end{eqnarray}
The equations of motion are now determined by Euler-Lagrange equations:
\begin{eqnarray}
    \frac{d}{dt}\Big(\frac{\partial \mathcal{L}_{eff}}{\partial \dot\theta} \Big) - \frac{\partial \mathcal{L}_{eff}}{\partial \theta} + \frac{\partial \mathcal{R}}{\partial \dot\theta} = 0,   \label{Eulag1}\\
    \frac{d}{dt}\Big(\frac{\partial \mathcal{L}_{eff}}{\partial \dot\phi} \Big) - \frac{\partial \mathcal{L}_{eff}}{\partial \phi} + \frac{\partial \mathcal{R}}{\partial \dot\phi} = 0. \label{Eulag2} 
\end{eqnarray}
We solve these equations and linearize them around the equilibrium (ground-state) equilibrium values $\theta_0$ and $\phi_0$, which are found by minimizing (\ref{phi}) yielding $\phi_0 = \pi/2$ and $\theta_0$ depending on the ratio of external field to anisotropy (i.e. when $H_{ext} = 0$ we have $\theta_0 = \pi/2$ while $\theta_0 = 0$ when $H_{ext} \gg H_{anis}$). Linearizing around these values, i.e. $\theta = \theta_0 + \theta_l$ and $\phi = \phi_0 + \phi_l$ with $\theta_l, \phi_l \ll 1$, the first equation (\ref{Eulag1}) gives:
\begin{eqnarray}
\ddot{\theta}_l + \frac{\alpha\mathcal{M}\overline{\gamma}}{\chi_\perp}\dot{\theta}_l + \Big(-\overline{\gamma}^2H^2\cos2\theta_0 + \frac{m\overline{\gamma}^2H}{\chi_\perp}\cos\theta_0 - \frac{2K_U\overline{\gamma}^2}{\chi_\perp} \cos2\theta_0\Big)\theta_l 
+ \Big(\overline{\gamma} H \sin2\theta_0 - \frac{m\overline{\gamma}^2}{\gamma_{ef}\chi_\perp}\sin\theta_0\Big)\dot\phi_l \nonumber\\ + \Big(-\overline{\gamma}^2Hh_y\cos2\theta_0 + \frac{m\overline{\gamma}^2h_y}{\chi_\perp}\cos\theta_0\Big)\phi_l 
= - \overline{\gamma}\dot h_y + \frac{\overline{\gamma}^2H^2}{2}\sin2\theta_0 - \frac{m\overline{\gamma}^2H}{\chi_\perp}\sin\theta_0 + \frac{\overline{\gamma}^2K_U}{\chi_\perp}\sin2\theta_0.
\label{theta_eqn}
\end{eqnarray}
where we introduced the notation $\chi_\perp \equiv \frac{\mathcal{M}^2}{\delta}$. Similarly for the second Euler-Lagrange equation (\ref{Eulag2}):
\begin{eqnarray}
\ddot\phi_l + \frac{\alpha \mathcal{M}\overline{\gamma}}{\chi_\perp}\dot\phi_l + \phi_l\Big(-\overline{\gamma}\dot h_y\cot\theta_0+ \overline{\gamma}^2h_y^2+\frac{2K_U\overline{\gamma}^2}{\chi_\perp}\Big) +\dot\theta_l\Big(-2\overline{\gamma} H \cot\theta_0 + \frac{m\overline{\gamma}^2}{\gamma_{ef}\chi_\perp\sin\theta_0}\Big) \nonumber \ \ \ \ \ \ \ \ \ \ \ \ \ \ \ \ \ \ \ \ \  \\
+ \ \theta_l\Big(-2\overline{\gamma}\dot h_x\cot\theta_0 + \overline{\gamma}^2 Hh_y(1-\cot^2\theta_0)+ \frac{\overline{\gamma}^2mh_y}{\chi_\perp}\frac{\cos\theta_0}{\sin^2\theta_0}\Big)
= \overline{\gamma} \dot h_x + \overline{\gamma}^2 Hh_y\cot\theta_0 - \frac{m\overline{\gamma}^2h_y}{\chi_\perp}\frac{1}{\sin\theta_0}. \label{motion2} 
\end{eqnarray}
These equations can be drastically simplified by noting that $\frac{1}{\chi_\perp}$ is proportional the the exchange constant and is therefore relatively large. Also the field derivative term $\gamma \dot{h}_i$ is strong, while terms proportional to $\sim \gamma h_{x,y}$ within brackets are driving terms proportional to the response and thus negligible. Equations (\ref{theta_eqn})-(\ref{motion2}) are then given in approximation:
\begin{multline}
\label{thetasimple}
\ddot{\theta}_l + \frac{\alpha\mathcal{M}\overline{\gamma}}{\chi_\perp}\dot{\theta}_l +  \Big(-\overline{\gamma}^2H^2\cos2\theta_0+\frac{m\overline{\gamma}^2H}{\chi_\perp}\cos\theta_0 - \frac{2K_U\overline{\gamma}^2}{\chi_\perp} \cos2\theta_0\Big)\theta_l +\Big(\overline{\gamma} H \sin2\theta_0 - \frac{m\overline{\gamma}^2}{\gamma_{ef}\chi_\perp}\sin\theta_0\Big)\dot\phi_l \\ = - \overline{\gamma}\dot h_y - \frac{m\overline{\gamma}^2H}{\chi_\perp}\sin\theta_0 + \frac{\overline{\gamma}^2K_U}{\chi_\perp}\sin2\theta_0,
\end{multline}
\begin{equation}
    \ddot\phi_l + \frac{\alpha \mathcal{M}\overline{\gamma}}{\chi_\perp}\dot\phi_l +\frac{2K_U\overline{\gamma}^2}{\chi_\perp}\phi_l + \Big(-2\overline{\gamma} H \cot\theta_0 + \frac{m\overline{\gamma}^2}{\gamma_{ef}\chi_\perp\sin\theta_0}\Big)\dot\theta_l = \overline{\gamma} \dot h_x + \overline{\gamma}^2 Hh_y\cot\theta_0 - \frac{m\overline{\gamma}^2h_y}{\chi_\perp}\frac{1}{\sin\theta_0}.
\end{equation}

The large field derivatives of the THz field $\dot h_{x,y}$ appear as a dominant driving force in these equations of motion. Interestingly, only the $y$-component $\dot h_{y}$ appears in the equation of motion for $\theta_l$, which we use here to understand the qualitative difference in dependencies on THz pump polarization assuming the field-derivative driving force is dominant.

In the experiment we saturate the magnetization with the external field at a small angle $\theta_0\approx 0$ (thus $\theta_0 \approx \pi$ for $\mathbf{-H}_{ext}$). Given that $\phi_0 = \pi/2$, we have that the modulations in the magnetization $z$-component are $M_z(t) = M\sin\phi\sin\theta \sim \pm \theta_l$ for $\pm\mathbf{H}_{ext}$ external field. The experiment reveals we are only sensitive to $M_z(t)$, so the detectable Faraday rotation modulations should also be proportional $\sim \pm \theta_l(t)$. When $\mathbf{H}_{THz} \perp \mathbf{M}$, $\dot h_x = h_x \neq 0$ and we have a strong non-zero driving force in \eqref{thetasimple}, it explains why we see immediate strong oscillations in $M_z$. Because the driving term has the same sign for both external field polarities $\pm \mathbf{H}_{ext}$, the forced oscillations must be sensitive to the polarity of the external magnetic field $\ddot M_z(t = 0) = \frac{d^2}{dt^2} \sin\big(\theta_0 + \theta_l(t)\big)\Bigr\rvert_{t = 0} \sim \pm \ddot\theta_l(t=0) \sim \mp  \gamma \dot h_y$ (as $\theta_0 = 0,$ $\pi$ for $\pm \mathbf{H}_{ext}$) i.e. this is an $\mathbf{H}$-odd effect. After the THz pulse has left the sample, the system of equations resembles those for a harmonic oscillator in $2$D, meaning the subsequent free oscillations will have opposite phases in the cases of opposite polarities of the external magnetic field.

Meanwhile by a similar argument, it is clear why a strong response is absent in $M_z$ when $\mathbf{H}_{THz} \parallel \mathbf{M}_{\parallel}$ ($\dot h_y = h_y = 0$) as in this case only the equation of motion for the in-plane dynamics $\phi_l(t)$ (Eq. \eqref{motion2}), to which we are not sensitive, has an initial non-zero driving force $\gamma \dot h_x$ while $\theta_l(t)$ does not. Detectable oscillations in $\theta_l(t)$ are instead only driven by cross-terms like $-\frac{m\overline{\gamma}^2}{\gamma_{ef}\chi_\perp}\sin\theta_0\dot\phi_l$ (Eq. \eqref{thetasimple}). Here it is important that the ground state $\theta_0$ is not exactly equal to $0$ and $\pi$, i.e. $\sin(\theta_0) = \pm \rho$ for $\pm \mathbf{H}_{ext}$ with $\rho > 0$ small constant (due to experimental canting of external field, otherwise no dynamics in this case is observed as was also seen in the experiment and simulations). Thus for opposite external field polarities, the driving force in $\theta_l$ has opposite sign $\mp \frac{m\gamma}{\chi_\perp}\rho\dot\phi_l $, contrary to the previous case. This means that subsequent oscillations are now expected to be even with $\mathbf{H}_{ext}$, in accordance with what was seen experimentally. Because these field-even oscillations are a secondary result from primary in-plane oscillations $\phi(t)$, it also explains why the observed effects are relatively weak when $\mathbf{H}_{THz} \parallel \mathbf{M}_{\parallel}$ compared to $\mathbf{H}_{THz} \perp \mathbf{M}$.

The eigenfrequencies in the article have been found by solving the coupled set of equations:
\begin{eqnarray}
    \label{motiontheta}
    \ddot{\theta}_l +  \frac{2K_U\overline{\gamma}^2}{\chi_\perp}\theta_l - \frac{m\overline{\gamma}^2}{\gamma_{ef}\chi_\perp}\dot\phi_l  &=& 0, \\
    \ddot\phi_l +\frac{2K_U\overline{\gamma}^2}{\chi_\perp}\phi_l  + \frac{m\overline{\gamma}^2}{\gamma_{ef}\chi_\perp}\dot\theta_l &=& 0.
\end{eqnarray}
Assuming $\theta_l, \phi_l \sim \exp(i\omega t)$ the frequencies can be solved by the equation
\begin{equation}
    \begin{vmatrix}
    -\omega^2  + \omega_K^2 & -i \omega \omega_{ex} \\
    i \omega \omega_{ex} &  -\omega^2  + \omega_{K}^2
    \end{vmatrix} = 0,
\end{equation}
where $\omega_K^2 = 2K_U\overline{\gamma}^2/\chi_\perp $ and $\omega_{ex} = m\overline{\gamma}^2/\gamma_{ef}\chi_\perp$
in the case of weak anisotropy $\omega_{ex} \gg \omega_K$, and we obtain:
\begin{equation}
    \omega = \pm \frac{\omega_{ex}}{2} \pm \sqrt{\frac{\omega_{ex}^2}{4} + \omega_K^2} \approx \pm \frac{\omega_{ex}}{2} \pm (\frac{\omega_{ex}}{2} + \omega_K^2/\omega_{ex}).
\end{equation}
Thus we obtain two approximate absolute frequencies:
\begin{eqnarray}
\omega_1 &=& \omega_K^2/\omega_{ex} = \gamma_{ef} \frac{2K_U}{m},\\
\omega_2 &\approx& \omega_{ex} = \frac{m\overline{\gamma}^2}{\gamma_{ef}\chi_\perp} \approx |\Lambda|(|\gamma_{R}|M_{Fe} - |\gamma_{Fe}|M_{R})
\end{eqnarray}
where in the last approximation we used that $M_{Fe}M_{R} \approx (M_{Fe} + M_R)^2/4$ and $(M_{Fe}/\gamma_{Fe} + M_{R}/\gamma_{R})^2 \approx  (M_{Fe}+M_{R})^2/\gamma_{Fe}\gamma_R$ to recover the approximate Kaplan-Kittel expression of exchange resonance.
\bibliographystyle{ieeetr}
\bibliography{supplement}